\documentclass[aps,prb,twocolumn,floatfix,floats,showpacs,superscriptaddress,raggedbottom]{revtex4}
\usepackage{graphicx,latexsym}
\usepackage{dcolumn}
\usepackage{amsmath}

\begin{document}

\title{Magnetotransport in a double quantum wire: \\ Modeling using a
    scattering formalism built on the Lippmann-Schwinger equation}

\author{Vidar Gudmundsson}
\affiliation{Science Institute, University of Iceland,
        Dunhaga 3, IS-107 Reykjavik, Iceland}

\author{Chi-Shung Tang}
\affiliation{Physics Division, National Center for Theoretical
        Sciences, P.O.\ Box 2-131, Hsinchu 30013, Taiwan}

%

\begin{abstract}
We model electronic transport through a double quantum wire
in an external homogeneous perpendicular magnetic field using
a scattering formalism built on the Lippmann-Schwinger equation.
In the scattering region a window is opened between the parallel
wires allowing for inter- and intra-wire scattering processes. 
Due to the parity breaking of the magnetic field the ensuing subband 
energy spectrum of the double wire system with its regimes of hole- 
and electron-like propagating modes leads to a more structure rich 
conductance as a function of the energy of the incoming waves 
than is seen in a single parabolically confined quantum wire.
The more complex structure of the evanescent modes of the system
also leaves its marks on the conductance. 
\end{abstract}



\maketitle

%
%

\section{Introduction}

We investigate coherent electronic transport in a
ballistic double-quantum-wire (DQW) system. The electronic
propagating modes in the two wires are allowed to be coupled through
a window between the separated wires. Such a wave-function mixing
can be enhanced in the presence of a perpendicular parity breaking
magnetic field. 

Earlier, discrete models of a single quantum wire with embedded island-like
obstacles have been investigated by Gu et al.\ using a transfer matrix
technique to describe the transport channels around the islands as two coupled
chains.\cite{Gu92:13274} A similar continuous model has been studied
by Tang et al.\ in order to identify coupling modes between the
channels.\cite{Tang05:195331} The question still remains how would
the transport be in a double wire system where the electron waves 
enter the scattering region already in eigenstates of the double wire. 

In the presence of the perpendicular magnetic field the energy spectrum 
of pure double quantum wire homogeneous in the long direction has been
investigated by Barbosa and Butcher pointing out the especially complex structure
of the evanescent states of the system.\cite{Barbosa97:325}
The properties of the energy spectrum of a double wire system in a magnetic
field has also been addressed by Korepov and Liberman who calculate the 
effects of a single delta impurity or disorder and boundary roughness in
a weakly coupled system of quantum wires.\cite{Korepov99:13770,Korepov02:92}

We have used the methodology of Gurvitz\cite{Gurvitz95:7123} to solve the
Lippmann-Schwinger equation for a single parabolic quantum wire in an external
perpendicular magnetic field in a momentum-coordinate space, transforming
the 2D-equation into a coupled set of 1D integral equations, one for each
incoming transport mode.\cite{Gudmundsson05:BT}
In this paper we use the eigenfunctions of the parabolically confined single
quantum wire in a magnetic field as a basis to expand the more complex 
wave functions of a double wire system in, and thus again arrive at a 
coupled set of 1D integral equations for the $T$-matrix that we then 
subsequently use to extract information about the conductance from
according to the Landauer-B{\"u}ttiker formalism, or the
electronic probability density of the system.\cite{Gudmundsson05:BT}  
We are thus able to handle
a system of weakly or strongly coupled quantum wires of a wide variety of
shapes.

%
\section{Model}
\begin{figure}[htbq]
      \includegraphics[width=0.45\textwidth,angle=0]{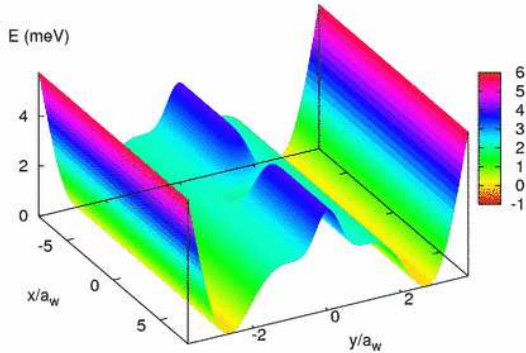}
      \caption{(Color online) A section of the double quantum wire 
      showing the scattering potential, the window between the wires.
      The color scale on the right shows the potential height in meV.
      The effective magnetic confinement length $a_w = 33.7$ nm,
      and other parameters are in the text.}
      \label{Wire}
\end{figure}

We consider a quantum wire extended homogeneously in the $x$-direction,
but the confinement in the $y$-direction is governed by
\begin{equation}
      V_{\rm conf}(y) = \frac{1}{2}m^*\Omega_0^2 y^2 + 
      V_{\delta}(y)D + E_{00},
\label{V_conf}
\end{equation}
with a symmetric deviation from the parabolic confinement
\begin{align}
      V_{\delta}(y)=&- V_{d_1} \exp{(-\beta_1 (y-y_0)^2)}\nonumber\\
                    &+ V_{d_0} \exp{(-\beta_0  y^2        )}\nonumber\\
                    &- V_{d_1} \exp{(-\beta_1(y+y_0)^2)}.
\label{V_delta}
\end{align}
The dimensionless parameter $D$ takes values in the 
interval $[0,1]$ and controls the overall strength
of the modulation of the double quantum wire.   
The scattering potential is of the type
\begin{equation}
      V_{\rm sc}(x,y) = V_0\exp{(-\alpha x^2-\beta_0 y^2)},
\label{Vsc}
\end{equation}
and with the choice $V_0 = -V_{d_0}$ can be made to represent a
window between the parallel wires as is shown in Fig.\ \ref{Wire}
for the parameters: $V_{d_1} = 6.0$ meV, $V_{d_0} = -V_0 = 2.0$
meV, $\hbar\Omega_0 = 1.0$ meV, $\beta_0 = 4.0\times 10^{-3}$
nm$^{-2}$, $\alpha = 0.2\times 10^{-3}$ nm$^{-2}$, $\beta_1 =
0.7\times 10^{-3}$ nm$^{-2}$, and $y_0 = 100$ nm. $E_{00}=2.0$ meV
is a convenient zero point to avoid negative energy in the potential.
In the case of a parabolically confined quantum wire in perpendicular
magnetic field we have used the approach of Gurvitz using a mixed
momentum-coordinate representation\cite{Gurvitz95:7123} of the wave
functions
\begin{equation}
      \Psi_E(q,y) = \sum_n \varphi_n(q)\phi_n(q,y),
\end{equation}
in order to reduce the two-dimensional Lippmann-Schwinger equation into
a coupled set of one-dimensional integral equations for the
T-matrix,\cite{Gudmundsson05:BT} where $\phi_n$ is an eigenfunction
for the parabolic wire in the presence of the magnetic field. 
Now we use the expansion
\begin{equation}
      \Psi_E(q,y) = \sum_n \varphi_n(q)\Phi_n(q,y),
\end{equation}
with the eigenfunctions $\Phi_n(q,y)$ of the new double wire confinement
Eq.\ (\ref{V_conf}) expanded in the basis for the parabolic confinement
\begin{equation}
      \Phi_n(q,y) = \sum_n c_{nm}(q)\phi_m(q,y).
\end{equation}
The coefficients $c_{nm}(q)$ are found by diagonalizing the corresponding
Hamiltonian in each $q$-subspace separately.
In this approach the momentum or the Fourier variable $q$ takes over the
role of the ``center-coordinate'' $y_0$ but loosens up the strict connection
between the center-coordinate and the momentum of the electron required by
the magnetic field. The Lippmann-Schwinger equation is separable in the
$(q,y)$-space but is not so in the $(x,y)$-space due to the Lorentz
force.\cite{Gurvitz95:7123,Gudmundsson05:BT}

This diagonalization of the Hamiltonian for the pure wire without an
embedded scattering center supplies the energy bands $E_n(q)$ for the
propagating states (when $q$ is real) in the asymptotic regions
of the double wire
\begin{equation}
      E_n(q) = E_n^0+\epsilon (n,q) + \frac{(qa_w)^2}{2}
      \frac{(\hbar\Omega_0)^2}{\hbar\Omega_w}
\label{E_nq}
\end{equation}
with $E_n^0 = \hbar\Omega_w(n+1/2)$. The effective confinement
frequency $\Omega_w$ and the effective magnetic length $a_w$ are
connected through
\begin{equation}
      a_w^2 = \frac{\hbar}{m^*}\frac{1}{\Omega_w} =
      \frac{\hbar}{m^*}\frac{1}{\sqrt{\Omega_0^2+\omega_c^2}},
\label{a_w}
\end{equation}
with $\omega_c=eB/(m^*c)$ the cyclotron frequency. The energy
$\epsilon (n,q)$ is the correction to the spectrum of the parabolic
wire due to the deviation from the parabolic confinement
(\ref{V_delta}). Its dependence on the band index $n$ makes the
subbands generally not equidistant in energy. 

In the open quasi-one-dimensional system considered here we are studying elastic 
scattering of the electrons. The incoming and outgoing states
have thus the same definite energy $E$ in the asymptotic regions
of the wire. The Lippmann-Schwinger equation includes all virtual
energy-nonconserving transitions in the scattering region.
A virtual state can be a bound state, a propagating one, or a
quasi-bound state. Evanescent states are localized quasi-bound
states with finite life time. The subband structure of the 
energy spectrum of the wire embeds them into the energy continuum
making them essential in order to obtain the correct conductance
of the system, especially when the energy of the incoming electrons
is close to a subband bottom.  

In order to calculate the energy of the evanescent states 
(with $q$ imaginary) we can not use a procedure similar to 
the one we used to evaluate the energy spectrum of the propagating 
states. The evanescent states are not orthogonal and the resulting 
generalized eigenvalue
problem is almost singular except for $q\approx i0^+$. An
observation of the character of evanescent wave functions for the
parabolic confinement explains these difficulties. An evanescent
wave function is always centered around the middle of the wire, $y =
0$, and develops a high number of nodes as $q$ assumes a higher
imaginary value. On the other hand a propagating state has a wave
function that shifts along the $y$-direction as $q$ is changed and
its number of nodes correlates to the quantum number $n$. Our
solution is thus to expand the evanescent states in terms
of the eigenfunctions of the parabolic wire at zero magnetic field,
but using in them the effective magnetic length for $B\neq 0$
\begin{equation}
      \phi_n^0(y) = \frac{\exp{\left(-\frac{y^2}{2a_w}\right)}}{\sqrt{2^n\:\sqrt{\pi}
      \: n!\: a_w}}H_n\left(\frac{y}{a_w}\right) .
\label{phi_0}
\end{equation}
The complete basis (\ref{phi_0})
thus used for the expansion of the evanescent states is composed of wave functions
that are all centered around $y = 0$ as are the evanescent wave functions for
a parabolically confined wire. In the numerical calculations the truncation of
the bases $\{\phi_n^0(y)\}$ and $\{\phi_n^0(q,y)\}$ is not performed for the same $n$.
The character of the evanescent wave functions requires a much larger basis for
their expansion.

With the basis (\ref{phi_0}) and an imaginary $q$ to find the energy spectrum
of the evanescent states the effective Hamiltonian matrix with the  
confinement potential (\ref{V_conf}) but without the scattering potential
(\ref{Vsc}) we need to diagonalize
is non-Hermitian as long as $\omega_c \neq 0$. The evanescent modes have real
eigenvalues when they are not degenerate, but they have an imaginary part when
they are degenerate. We assume this reflects the fact that the wave vectors
of the evanescent modes can be complex when the states are calculated explicitly
without using the basis (\ref{phi_0}) we use.\cite{Korepov02:92}
We find the real part of our energy spectrum for the evanescent modes to be
consistent with the results of Barbosa and Butcher\cite{Barbosa97:325}
and Korepov and Libermann\cite{Korepov02:92} for nonparabolic confinement.
We thus discard the imaginary part of the energy spectrum for the
evanescent modes found by our method.

The electron scattering in the wire is elastic. For the parabolic wire this means
that an electron with energy $E$ in the asymptotic free region would have
a subband momentum $\pm k_n(E)$ if it could propagate in the subband with index $n$.
The deviation from the parabolic confinement requires the subband momentum
in subband $n$ to be found from the condition
\begin{equation}
      \left[k_n(E)a_w\right]^2 = 2\left[E-E_n^0-\epsilon (n,q)\right]
      \frac{\hbar\Omega_w}{(\hbar\Omega_0 )^2}.
\label{k_n}
\end{equation}
For the more complicated confinement assumed here each subband can support more
than one or two propagating states. These different transport modes correspond to the
poles, $q_{n_i}$, of the scattering Green function
\begin{equation}
      G_E^n(q) = \frac{1}{(k_n(E)a_w)^2-(qa_w)^2+i0^+},
\label{G_nE}
\end{equation}
and are located where the incoming energy $E$ intersects the energy subband $n$.
The index $i$ labels the modes in subband $n$ in the direction
of increasing $q$. 
The energy spectrum of the evanescent states enters the Green function
for the subbands with no propagating modes present, i.e.\ for subbands
with threshold energy higher than $E$.\cite{Bagwell90,Gurvitz93}

The resulting coupled one-dimensional Lippmann-Schwinger equations
for the transport modes supply the T-matrix and the probability
amplitudes for transmission in mode $n_i$ with momentum $\hbar k_{n_i}$ if
the in-state is in mode $m_j$ with momentum $\hbar k_{m_j}$
\begin{align}
      t_{{n_i}{m_j}}(E) = \delta_{{n_i}{m_j}} -&{\ } 
      \frac{i\sqrt{(k_{m_j}/k_{n_i})}}{2(k_{m_j}a_w)}
      \left(\frac{\hbar\Omega_0}{\hbar\Omega_w} \right)^2\nonumber\\
      &\times {\tilde T}_{{n_i}{m_j}}(k_{n_i},k_{m_j}).
\label{t(E)}
\end{align}
The conductance is then according to the Landauer-B{\"u}ttiker formalism
defined as
\begin{equation}
      G(E) = \frac{2e^2}{h}{\rm Tr}[ {\bf t}^{\dagger}(E){\bf t}(E)] ,
\label{G}
\end{equation}
where ${\bf t}$ is evaluated at the Fermi energy.
%

\section{Results and discussion}
In order to understand the conductance of the double wire system we
can correlate it to the energy spectrum in the asymptotic region far
away form the scattering region in the middle of the wire. To further
clarify the conductance for a certain incoming energy of the electrons
we will seek information from the electron probability density of the
various modes active at that point. We assume GaAs parameters here
with $m^*=0.067m_e$.

\begin{figure}[htbq]
      \includegraphics[width=0.42\textwidth,angle=0]{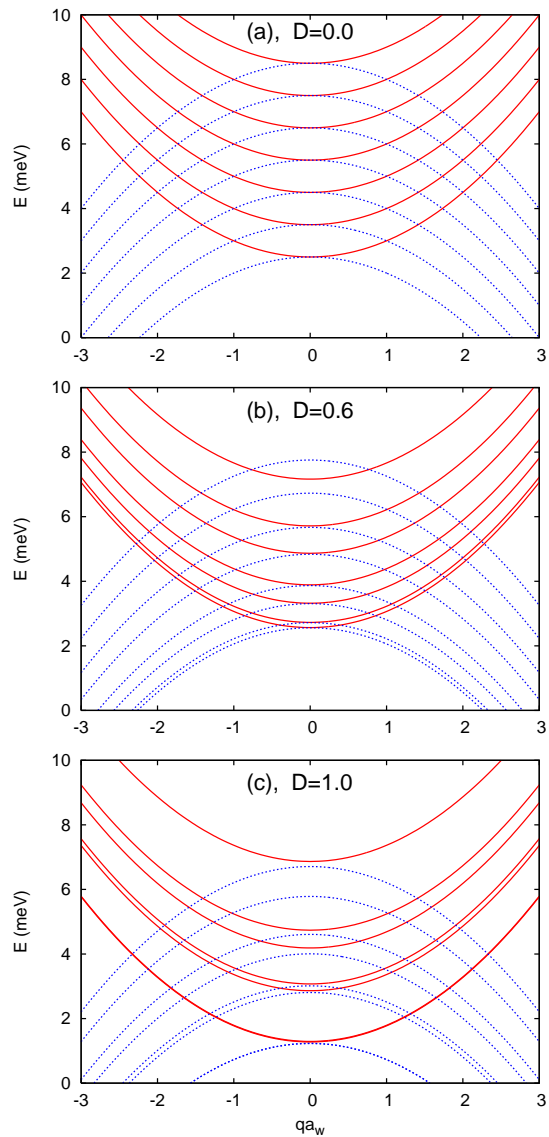} 
      \caption{(Color online) The energy spectrum of the propagating 
               electronic states (solid red) vs.\ the Fourier
               parameter $q$, and the energy spectrum of the evanescent
               states (dotted blue) vs.\ $iq$, 
               in the asymptotically free regions of the quantum wire.
               $D$ is an overall multiplicative factor to the 
               deviation $V_{\delta}(y)$ from the parabolic confinement.
               $B=0$ T.}
      \label{E_B00}
\end{figure}

\begin{figure}[htbq]
      \includegraphics[width=0.42\textwidth,angle=0]{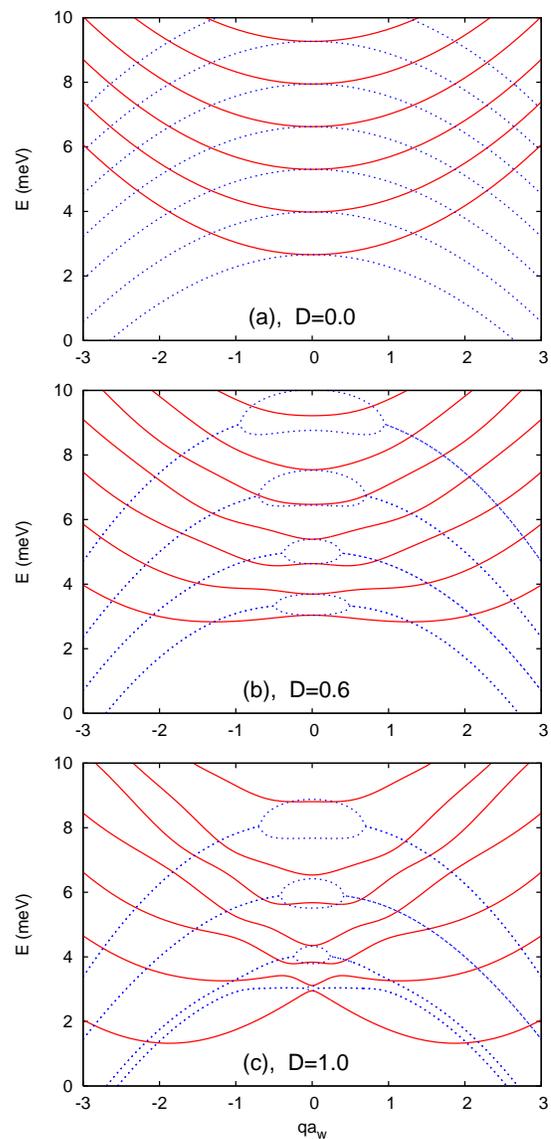}
      \caption{(Color online) The energy spectrum of the propagating 
               electronic states (solid red) vs.\ the Fourier
               parameter $q$, and the energy spectrum of the evanescent
               states (dotted blue) vs.\ $iq$, 
               in the asymptotically free regions of the quantum wire.
               $D$ is an overall multiplicative factor to the 
               deviation $V_{\delta}(y)$ from the parabolic confinement.
               $B=0.5$ T.}
      \label{E_B05}
\end{figure}

Figures \ref{E_B00} and \ref{E_B05} present some general 
features of the energy spectra at a vanishing magnetic field and at
a finite one. In Figure \ref{E_B00} the energy spectrum for 
$B=0$ is shown for three different strengths of modulation
of the wire shape. $D=0$ corresponds to a parabolic confinement
and the energy spectrum in Fig.\ \ref{E_B00}(a) displays the familiar
spectrum of a parabolic quantum wire in vanishing magnetic field.
We only show here 7 of each of the 9 propagating and 9 evanescent subbands
included in the present calculation.  
The shape of the wire determined by Eq.'s (\ref{V_conf}) and (\ref{V_delta}) 
is always symmetric around it's center line $y=0$ requiring the wave functions
to be either symmetric or antisymmetric in the case of no magnetic field.
The resulting energy subbands are parabolic and have their bottom (or top in case of 
the evanescent ones) at $q=0$ but are not equidistant any more since the deviation 
from the parabolic confinement (\ref{V_delta}) shifts the subbands to a different extend
depending on their parity. For $D=0$ in Fig.\ \ref{E_B00}(a) the energy gap between the
two lowest subbands $\Delta_{01}=\hbar\Omega_w=1.0$ meV at $q=0$.
In Fig.\ \ref{E_B00}(b) the lowest propagating subband representing
symmetric wave functions is shifted up in energy very close to the next subband of
antisymmetric wave functions and $\Delta_{01}=0.16$ meV. 
In Fig.\ \ref{E_B00}(c) the stronger parabolic
deviation causes a near accidental degeneration between these two lowest subbands
and higher subbands appear to come in pairs, $\Delta_{01}=0.017$ meV.  

The magnetic field changes the character of the energy spectrum as can be seen in
Fig.\ \ref{E_B05} by destroying the parity of the wave functions. 
Physically this can be related to the action of the Lorentz force that pushes the
electrons away from the center of the wire in ratio to their momentum along the
wire. 
In a parabolic wire only the energy separation of the subbands is renormalized
by the magnetic field (Fig.\ \ref{E_B05}(a)) with $\Delta_{01}=1.32$ meV, 
but in a double wire the 
minimum of a propagating energy subband does not have to be at $q=0$.
The evanescent subbands assume a complex form that has been detailed by
Barbosa and Butcher.\cite{Barbosa97:325} In Figure \ref{E_B05}(b) 
with $\Delta_{01}=0.66$ meV and in subfigure \ref{E_B05}(c) we
notice that states with a finite ``subband momentum'' $q$ can be off-centered by the
Lorentz force enough to occupy mostly only one of the parallel wires and
thus acquiring less potential energy leading to a minimum total energy at
$q\neq 0$. In addition, we see in  Figure \ref{E_B05}(c) that the value
$D=1$ leads to energy subbands formed by a strong admixture of several 
subbands for the parabolic wire, $\Delta_{01}=0.16$ meV.  

We now turn to analysis of the conduction properties of the double wire
system in an external magnetic field with the help of the energy spectra
for the asymptotic regions of the wire. The scattering potential (\ref{Vsc})
creates a shallow window between the parallel wires as is displayed in 
Fig.\ \ref{Wire}. 
In the case of an ideal parabolic wire the conductance increased monotonically 
step wise as a function of $E$, with a new step appearing at the onset of each
new propagating subband. For an ideal double wire system the steps do not increase
monotonically with increasing $E$, but instead reflect the much richer subband structure
of the system. 

\begin{figure}[htbq]
      \includegraphics[width=0.45\textwidth,angle=0]{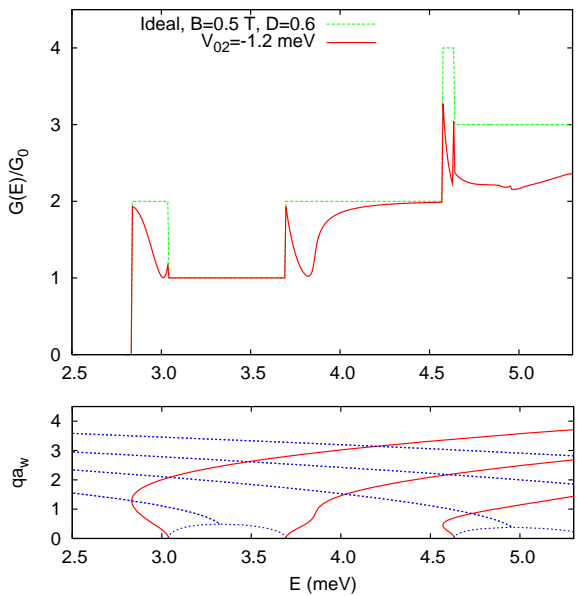}
      \caption{(Color online) (Upper panel) The conductance of an ideal 
               weakly ($D=0.6$) modulated double quantum wire without
               a window (dashed green), and with a window between
               the wires (solid red).
               (Lower panel) The energy spectrum of the propagating 
               electronic states (solid red) vs.\ the Fourier
               parameter $q$, and the energy spectrum of the evanescent
               states (dotted blue) vs.\ $iq$, 
               in the asymptotically free regions of the quantum wire.
               $B=0.5$ T, $\alpha = 0.0002$ nm$^{-2}$, and $V_0 = -1.2$ meV.}
      \label{GE_D06}
\end{figure}

Figure \ref{GE_D06} presents the conductance for the system
with the modulation of the parabolic deviation of the wire shape $D=0.6$
and the same reduction applies to the scattering potential (\ref{Vsc}).
Shortly after the energy of the incoming electrons hits the lowest 
subband bottom the conductance jumps to $2G_0$ indicating two propagating
modes. The two modes remain active with increasing in-energy $E$ until 
the local maximum at $q=0$ is surpassed at $E=3.03$ meV and the conductance
falls to $G_0$ as only one incoming mode is active. 
Between the global minimum and the local maximum of the subband at $q=0$
the conductance of the system goes through a minimum close to the point
of inflection separating the states of the subband that are hole-like 
close to $q\approx 0$ and electron like nearer the global minimum. 
For $E$ below the inflection point the incoming electron states travel
very close to the center barrier between the wires, and in addition the
Lorentz force pushes them towards the barrier. In the scattering region,
in the window, the Lorentz force manages (if the window is long enough
compared with the magnetic length) to push the electron wave partially 
into the barrier region. Part of the wave is thus reflected and a part
tunnels within the central barrier before emerging again as a traveling
wave close to the central barrier. The tunneling distance can be quite
considerable as we shall show in a figure in combination with the 
discussion about the transport in a wire with a stronger double wire
modulation ($D=1$). The hole-like states with $E$ higher than the 
inflection point do experience the Lorentz force that pulls them
away from the barrier. In the window region they are thus less likely
to be reflected and the conductance rises again. The effectiveness of
Lorentz force to shift the center of the states depends not only on their
respective kinetic energy, but also on their effective mass which is
singular at the point of inflection.  

This distinction in the behavior of hole- or electron-like states can
not be observed in a parabolic single quantum wire and in a double wire
system the magnetic field perpendicular to the wire is essential.   

In the semi-subband gap between $E=3.03$ and 3.7 meV the single outer
mode goes undisturbed through the wire system as the Lorentz force steers it
clear of the small window between the two parallel wires.    
Just above $E=3.7$ meV the second subband becomes active in the transport
and the total conductance of the system reaches $2G_0$. 
At the bottom of this subband there is no local maximum
at $q=0$ as in the first subband, but for a bit higher $q$ the
deviation from the parabolic shape of the system produces a point of inflection.
Again the states near the bottom of the second subband reside close to 
the small central barrier like those close to the bottom of the first
subband. The main difference now being that their wave function have 
a double hump structure
in the $y$-direction. The fine interplay of the Lorentz force and the 
window between the wires produces again a minimum of conductance around the
point of inflection in a similar manner as occurred in the first subband.  
With increasing $E$ the shape of the subband approaches a normal form for
electronic states and the conductance rises slowly to $2G_0$ since now states
in both subbands at higher values of $q$ steer clear of the window between the
wires. With the activation of the third subband we see a similar structure
in the conductance as observed at the beginning of the first subband, as
these two subbands have a similar shape for low $q$ values.

\begin{figure}[htbq]
      \includegraphics[width=0.45\textwidth,angle=0]{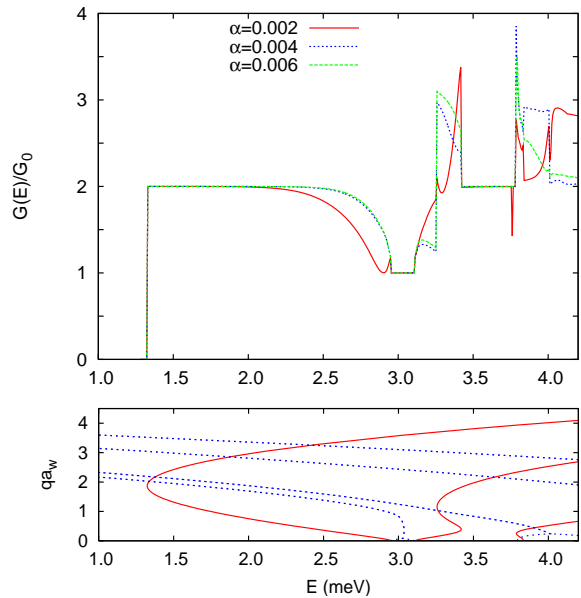}
      \caption{(Color online) (Upper panel) The conductance of a 
               double quantum wire with a window between the wires
               of different lengths indicated by the three different
               values of $\alpha$ used for the scattering potential
               (\ref{Vsc}).  
               (Lower panel) The energy spectrum of the propagating 
               electronic states (solid red) vs.\ the Fourier
               parameter $q$, and the energy spectrum of the evanescent
               states (dotted blue) vs.\ $iq$, 
               in the asymptotically free regions of the quantum wire.
               $B=0.5$ T, $D=1.0$, and $V_0 = -2.0$ meV.}
      \label{GEW_D10}
\end{figure}

\begin{figure}[htbq]
      \includegraphics[width=0.45\textwidth,angle=0]{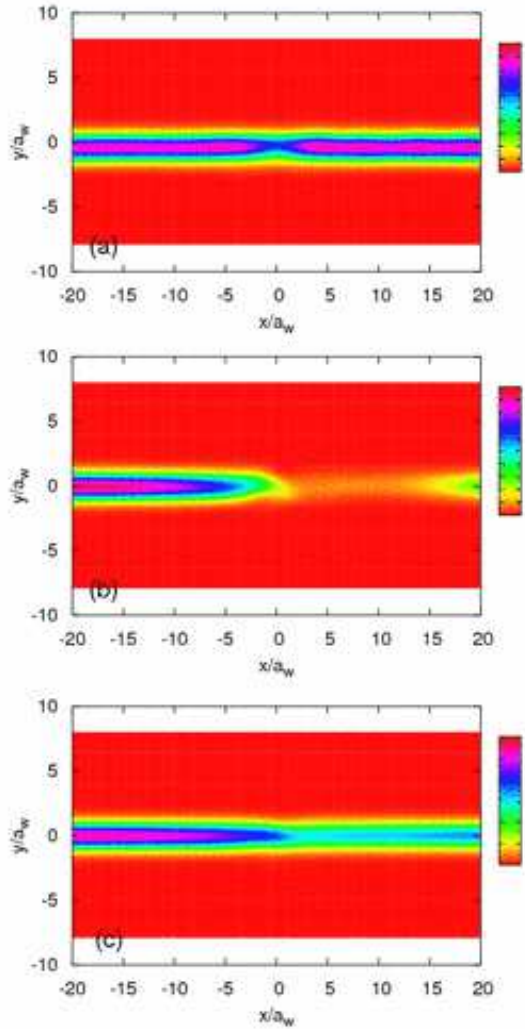}
      \caption{(Color online) The electron probability density 
               for the $n_i=0_2$ electron-like mode at input energy
               $E=2.155$ meV (a), for the mode at $E=2.902$ meV (b), 
               and for the hole-like mode at $E=2.942$ meV (c).
               $B=0.5$ T, $\alpha = 0.0002$ nm$^{-2}$, and $D=1.0$ 
               corresponding to the
               solid red curve in the upper panel of Fig.\ \ref{GEW_D10}.}
      \label{PD_e-h}
\end{figure}

The conductance for the stronger modulated double quantum wire 
with $D=1$ is shown in Fig.\ \ref{GEW_D10} for three different
lengths of the window between the parallel wires. The window length
is controlled by the parameter $\alpha$ introduced in the scattering
potential in eq.\ (\ref{Vsc}). The value $\alpha=0.0002$ nm$^{-2}$ was
used for the shape of the wire shown in Fig.\ \ref{Wire} and in the
conductance calculation for Fig.\ \ref{GE_D06}. Higher values of 
$\alpha$ lead to a shorter window.  
The conductance up to $E=2.95$ is in accordance with the results for the
lowest subband in Fig.\ \ref{GE_D06} caused by the fine tuned interplay of the
Lorentz force and the window size and the character of the transport states.
Here we use the opportunity to clarify the effects through the probability
distribution of the electrons. In Fig.\ \ref{PD_e-h} we display the probability
density for three different values of $E$ for the incoming state $n_i=0_2$,
in the double wire with the longer window with $\alpha =0.0002$ nm$^{-2}$, 
close to the conductance minimum at $E=2.902$ meV and the inflection point
of the subband. The other active incoming
transport mode ($n_i=0_4$) in this energy region passes the scattering region
unperturbed and is therefore not shown here. 

\begin{figure}[htbq]
      \includegraphics[width=0.45\textwidth,angle=0]{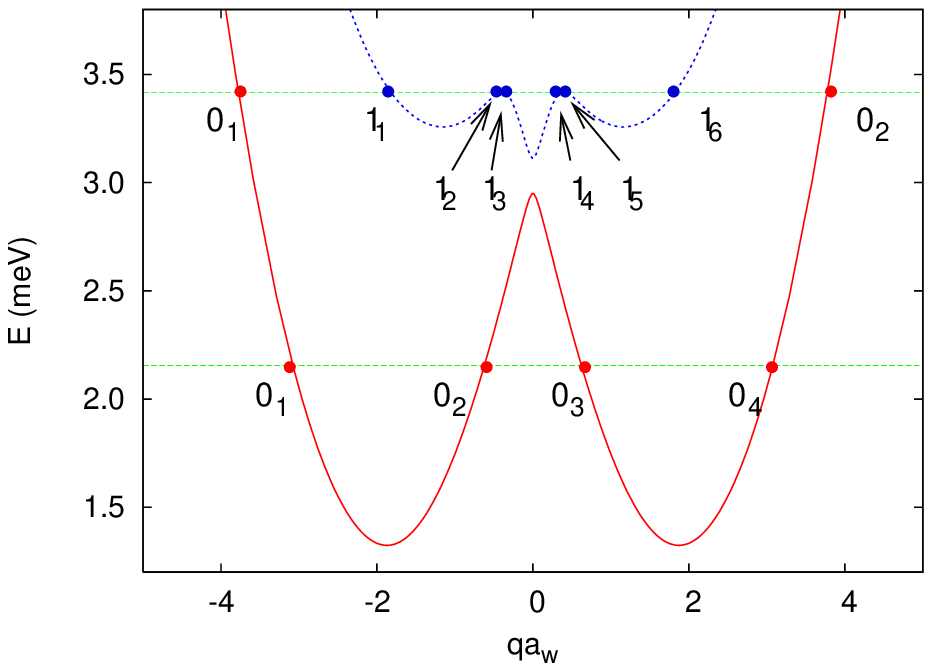}
      \caption{(Color online) The energy spectrum of the two lowest propagating 
               electronic subbands vs.\ the Fourier parameter $q$ in the asymptotically 
               free regions of the quantum wire. The two horizontal dashed green lines
               represent the energies $E=2.155$ meV and $E=3.417$ meV.
               The active transport modes are labeled for these two energies
               with the notation $n_i$, where $i$ is the number of the active
               mode in subband number $n$.
               $D=1$ and $B=0.5$ T. }
      \label{E_Info}
\end{figure}

To make the subsequent discussion of transport modes clearer we have in 
Fig.\ \ref{E_Info} labeled the active transport modes for two different
values of the incoming energy that are referred to later. 

Well below the inflection point the electron-like in-state at $E=2.155$ meV 
depicted in Fig.\ \ref{PD_e-h}(a) contributing $0.98$ G$_0$ to the total conductance  
enters the wire on the right side of the thin central barrier between the wires
and is only slightly perturbed in the window region. 
In Fig.\ \ref{PD_e-h}(b) corresponding to the conductance minimum at $E=2.902$ meV
the wave is deflected into the window region causing the largest part of it to
be backscattered and a small portion tunneling into the continuation of the 
central barrier on the other side of the window emerging as a traveling wave  
a good distance from the window. 
In case of a shorter window, $\alpha \geq 0.0004$ nm$^{-2}$, the electron wave 
interacts to a lesser extent with the window and no clear distinction
can be seen between the behavior of hole- or electron-like states.
The shorter window is getting small on the scale of the magnetic length
and can not effectively interfere with the wave with the help of the 
Lorentz force. The hole-like
in-mode seen in Fig.\ \ref{PD_e-h}(c) at $E=2.942$ meV avoids due to the opposite
directed Lorentz force the strong backscattering in the window region.
The accuracy of these results has been checked by increasing all relevant
grid or basis set sizes used in the numerical calculation. 
For the calculation here we have used 9 subbands, and 
divided the $q$-integration into three regions: $qa_w \in (0,3)$,
$(3,7)$, and $(7,10)$. In each region we apply a repeated 4-point Gauss
integration. For the calculation of the conductance we need 32 repetitions
in the first interval, 24 in the second, and 8 in the last one. 
This is then mirrored for the negative values of $q$. For the probability
density we double the density of the points, but use the same intervals.  

\begin{figure}[htbq]
      \includegraphics[width=0.45\textwidth,angle=0]{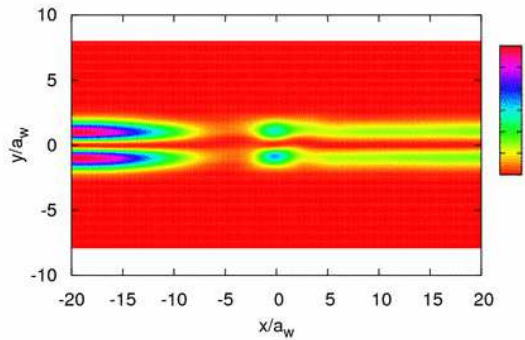}
      \caption{(Color online) The electron probability density 
               for the mode $n_i=1_2$ at input energy
               $E=3.212$ meV. 
               $B=0.5$ T, $\alpha = 0.0002$ nm$^{-2}$, and $D=1.0$ 
               corresponding to the
               solid red curve in the upper panel of Fig.\ \ref{GEW_D10}.}
      \label{Wf_E3p2124257_n4}
\end{figure}

In the semi-gap region $2.95<E<3.11$ meV only a single edge state away 
from the window region contributes $G_0$ to the conductance.
In the energy region $3.11<E<3.25$ meV when only one state in the lowest
part of the second subband contributes to the conduction there is a 
stronger backscattering for the smaller window sizes. We have checked
this region for even larger window with $\alpha = 0.0001$ nm$^{-2}$ and
find then the strongest backscattering for all our tested values of $\alpha$.
The reason for the suppressed backscattering at $\alpha = 0.0002$ nm$^{-2}$
is a formation of a resonance state in the window region for this particular
size of it. The electron probability density for this state $n_i=1_2$ is seen 
in Fig.\ \ref{Wf_E3p2124257_n4}.
The resonance is in some sense opposite to what happens at $E=2.902$ meV
in Fig.\ \ref{PD_e-h}(b), since here there is resonant forward scattering from the
window region but the backscattered wave tunnels initially into the barrier
left of the window.
For the same energy the $n_i=0_2$ state bypasses the scattering region
unperturbed. 
 
\begin{figure}[htbq]
      \includegraphics[width=0.41\textwidth,angle=0]{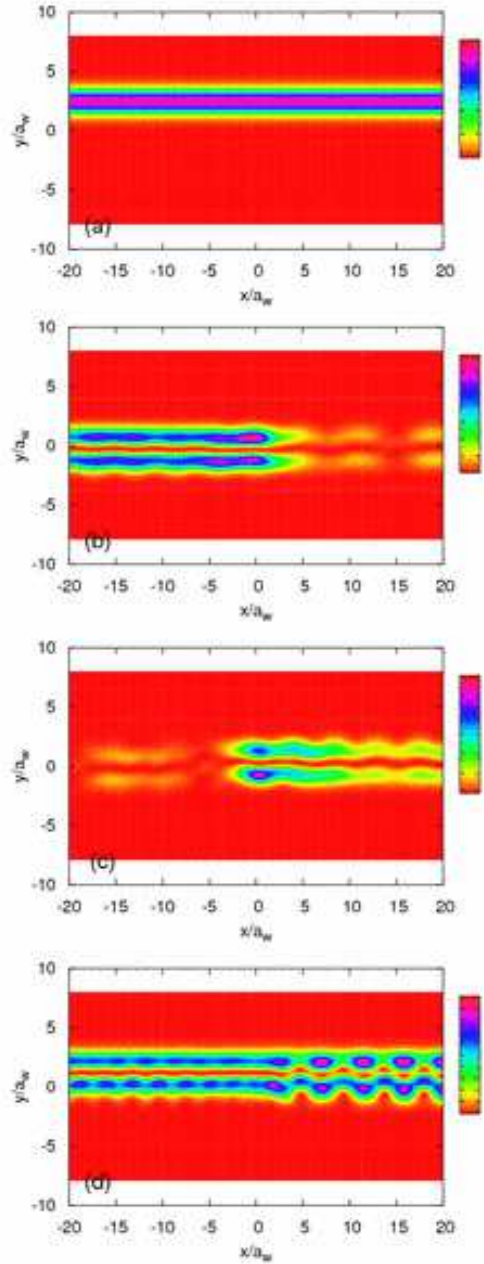}
      \caption{(Color online) The electron probability density at input
               energy $E=3.417$ meV for mode $n_i=0_2$ (a), $n_i=1_2$ (b), 
               $n_i=1_4$ (c), and $n_i=1_6$ (d). See Fig.\ \ref{E_Info}
               for the labeling of the modes.
               $B=0.5$ T, $\alpha = 0.0002$ nm$^{-2}$, and $D=1.0$ 
               corresponding to the
               solid red curve in the upper panel of Fig.\ \ref{GEW_D10}.}
      \label{PE_E3p4172154}
\end{figure}

In the energy region $3.25<E<3.417$ meV two more states in the second subband
can contribute to the conductance and again we have a complex superposition
of hole- and electron-like states interacting with the window leading to
a conductance minimum for the largest window, $\alpha = 0.0002$ nm$^{-2}$.
The electron probability density for $E=3.417$ meV in the peak at the endpoint of 
this interval is shown in Fig.\ \ref{PE_E3p4172154}. In subfigure \ref{PE_E3p4172154}(a) 
we see the incoming state in the lowest subband bypassing the window unperturbed,
but in Fig.\ \ref{PE_E3p4172154}(b) the $n_i=1_2$ hole-like mode undergoes
a inter-wire backscattering in the window and we see a faint wave continuing
through the wire. In subfigure \ref{PE_E3p4172154}(c) the other hole-like 
mode $n_i=1_4$ shows
a quasi-bound state in the window. The interference pattern
seen in these two hole-like states indicates a substantial mixing or scattering  
between them. Some particles coming into the system in the $n_i=1_2$ mode
leave the system in the $n_i=1_4$ mode. 
The last panel of Fig.\ \ref{PE_E3p4172154}(d) shows the only incoming
electron-like state $n_i=1_6$ in the second subband. Clearly the electrons
in this mode are not backscattered to any extent, but the interference
pattern again indicates interference with the two hole-like modes
propagating in the same subband. 

\begin{figure}[htbq]
      \includegraphics[width=0.45\textwidth,angle=0]{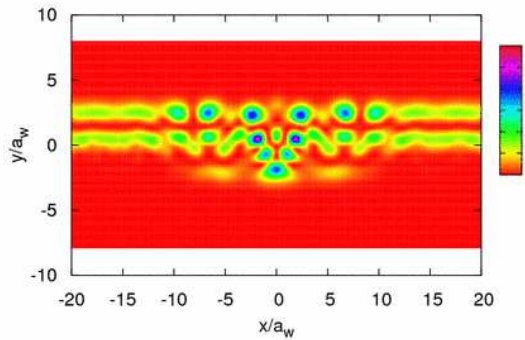}
      \caption{(Color online) The electron probability density at input
               energy $E=3.761$ meV for mode $n_i=1_2$.
               $B=0.5$ T, $\alpha = 0.0002$ nm$^{-2}$, and $D=1.0$ 
               corresponding to the
               solid red curve in the upper panel of Fig.\ \ref{GEW_D10}.}
      \label{Wf_E3p7607336_n4}
\end{figure}

In the semi-gap energy region $3.42<E<3.79$ meV the two active incoming
modes of the lowest two subbands contribute their full share to the 
conductance unperturbed by the central window, except for a narrow
conductance dip at $E=3.761$ meV where the $n_i=1_2$ mode interacts
strongly with an evanescent state at the bottom of the third subband.
The corresponding electron probability is seen in Fig.\ \ref{Wf_E3p7607336_n4}.
The breaking of the parity by the magnetic field makes this backscattering
through an evanescent state in the next higher subband possible.\cite{Gudmundsson05:BT}
The resonance state seen in Fig.\ \ref{Wf_E3p7607336_n4} is an example
of persistence of a wave function in an open system, or ``scarring of a
wave function'' discussed by Akis et al.\cite{Akis02:129}

In the narrow energy range $3.79<E<3.83$ meV two incoming modes are active in the
third subband and we have a structure in the conductance similar to what happened
around the bottom of the two lower subbands. But in the energy range $3.83<E<4.2$ meV
when only one incoming mode exists in the third subband we have a curious ``exchange''
of backscattering strength between the conduction curves for 
the longest window and the shorter ones.
The strength changes at a resonance caused by an evanescent state at $E=4.01$ meV.
The character of a close evanescent state changes at this energy, below it is electron-like
but above it is hole-like. For the larger window, $\alpha =0.0002$ nm$^{-2}$, the remnant
of the resonance enhances conductance above $E=4.01$ meV, and has opposite
effects for the larger windows.
%

%
\section{Summary}
In this work we have shown that magnetotransport in a lateral
double quantum wire system with a window between the parallel wires shows 
clear fingerprints of the complex subband structure of the system
caused by the parity breaking of the perpendicular magnetic field.
We have been able to specify regions in the conductance of the system
as a function of the incoming energy $E$ that reflect the underlying
electron- or hole-like states active in the transport process 
both among the propagating and the evanescent modes of the system. 
This connection of the transport features to the energy spectrum
was of course not possible in a single quantum wire with an island
in the center of the scattering region creating a short section of
a double quantum wire embedded in larger single wire.  

An expected simple yet nontrivial result of the complex subband structure for the
double wire system is the appearance of irregular steps in the conductance as a
function of the energy of the incoming electron wave compared to the
regular steps of the single quantum wire of parabolic or hard wall shape.

Here we have selected a shallow window as a particular scattering 
potential between parallel quantum wires of a particular shape
such that the states around subband bottoms are most affected by
inter-wire scattering processes. The numerical methods employed 
using analytical matrix elements allow for a wide variety of wire shape
and scattering potentials that can have more effects on edge-like
modes of the system.  


%
%
\begin{acknowledgments}
      The authors acknowledge the financial support by the Research
      and Instruments Funds of the Icelandic State,
      the Research Fund of the University of Iceland, and the
      National Science Council of Taiwan.
      C.S.T.\ is grateful to the computational facility supported by the
      National Center for High-performance Computing of Taiwan.
\end{acknowledgments}

%
%
\bibliographystyle{apsrev}

%
%
%
\end{document}